# CORE-Deblur: Parallel MRI Reconstruction by Deblurring Using Compressed Sensing

Efrat Shimron[1], Andrew G. Webb[2], Haim Azhari[1*]

[1] Biomedical Engineering Department, Technion - Israel Institute of Technology, Haifa, 3200003, Israel
[2] C.J. Gorter Center for High Field MRI Research, Department of Radiology, Leiden University Medical Center (LUMC), Leiden, The Netherlands
[*] Corresponding author: haim@bm.technion.ac.il

*Abstract* – **In this work we introduce a new method that combines Parallel MRI and Compressed Sensing (CS) for accelerated image reconstruction from subsampled k-space data. The method first computes a convolved image, which gives the convolution between a user-defined kernel and the unknown MR image, and then reconstructs the image by *CS-based image deblurring*, in which CS is applied for removing the inherent blur stemming from the convolution process. This method is hence termed CORE-Deblur. Retrospective subsampling experiments with data from a numerical brain phantom and in-vivo 7T brain scans showed that CORE-Deblur produced high-quality reconstructions, comparable to those of a conventional CS method, while reducing the number of iterations by a factor of 10 or more. The average Normalized Root Mean Square Error (NRMSE) obtained by CORE-Deblur for the in-vivo datasets was 0.016. CORE-Deblur also exhibited robustness regarding the chosen kernel and compatibility with various k-space subsampling schemes, ranging from regular to random. In summary, CORE-Deblur enables high quality reconstructions and reduction of the CS iterations number by 10-fold.**

## I. INTRODUCTION

Reducing MRI acquisition time is beneficial in terms of improved patient comfort, reduced motion artifacts, and increased system throughput. Scan time reduction can be achieved by sampling k-space with a rate lower than the Nyquist sampling rate; however, suitable reconstruction techniques are then needed for accurate image reconstruction. Parallel Imaging (PI) and Compressed Sensing (CS) are two well-established frameworks developed for this aim.

In PI, multiple coils simultaneously acquire the k-space data. The data acquired by each coil is weighted by an individual coil-specific sensitivity function, and the reconstruction process utilizes knowledge of the coil sensitivities and the sampling trajectory to compensate for k-space sub-sampling. PI methods such as SENSE [1] and GRAPPA [2] are well-established in a wide range of clinical implementations.

Compressed Sensing (CS) is a more recent framework for image reconstruction from sub-sampled k-space data. CS methods commonly exploit the data sparsity in a known transform domain, such as the wavelet domain, and reconstruct the image by solving a convex optimization problem. The application of CS to MRI was introduced by Lustig et al. [3] and has been widely explored since [4–8].

A large body of work has demonstrated that PI and CS can be combined for MRI scan time reduction. PI-CS methods have been applied to Cartesian [9] and non-Cartesian [10] imaging in a wide range of applications, including body imaging [11,12] and dynamic cardiac MRI [7]. However, despite the advantages offered by PI-CS techniques, currently the CS data reduction factors are typically only about two or three. One of the reasons is that CS ideally requires purely random k-space undersampling, which is not optimal for PI methods; for example, the popular commercially-implemented SENSE method is efficient and non-iterative when k-space is uniformly subsampled in a single direction, but requires iterative reconstruction for other sampling schemes [13].

Here, we introduce a new PI-CS reconstruction method, CORE-Deblur, which enables reconstructions from sub-sampled k-space data with a variety of different subsampling schemes. The CORE-Deblur method includes two steps: (i) computation of a *convolution image*, which describes the convolution between the target MR image and a known user-defined kernel, and (ii) a CS process, which is initiated from the convolution image and reconstructs the MR image. CORE-Deblur therefore introduces a novel concept: *reconstruction by*

*sparsity-based image deblurring*. Furthermore, the method is implemented with very few iterations, and enjoys the benefit of additional implicit regularization obtained by this small number of iterations, i.e. early stopping.

The main contributions of this paper are: (i) development of a new PI-CS reconstruction approach, (ii) introduction of the novel concept of applying CS for image deblurring, and (iii) demonstration of the proposed method using simulations and in-vivo brain data. (Some aspects of this work have been presented in a conference proceeding [14]).

## II. THEORY

*Overview*

CORE-Deblur is a method for image reconstruction from subsampled 2D Cartesian k-space data. The method requires three inputs: (i) multi-coil subsampled k-space data, (ii) estimations of the coils sensitivity maps, and (iii) a user-defined 1D kernel (see following explanation). As mentioned above, CORE-Deblur is applied in two steps:

1. Computation of the *convolution image* $f^{conv}(x, y)$, which describes the 1D row-wise convolution between the *unknown* MR image $f(x, y)$ and a *known* user-defined kernel $g(x)$. This step utilizes the recently introduced Convolution-based Reconstruction (CORE) technique [15].
2. Image reconstruction via a CS $l1$-minimization process that is initiated from $f^{conv}(x, y)$. This process removes the blurring that is inherent to the convolution image and recovers the final image $f(x, y)$. The CS process is implemented with an early stopping that brings in additional implicit regularization.

We now turn to a mathematical description of the proposed method.

*k-space subsampling*

As stated, CORE-Deblur is suitable for multi-coil 2D k-space data acquired on a Cartesian grid, with subsampling along one dimension (which is typically the phase encoding dimension) and full sampling along the other dimension [16]. The method also assumes that the individual coil sensitivities are known; they may be calibrated for example using a preliminary low-resolution scan.

*Required Inputs*

Let $f(x, y)$ be the unknown $N \times N$ image to be reconstructed and let $N_c$ be the number of coils used. The required inputs of CORE-Deblur are:

- The estimated sensitivity maps $C_{n_c}(x, y)$ for $n_c \in [1, N_c]$.
- A 1D user-defined kernel function $g(x)$ of length $N$.
- The acquired multi-coil k-space signals $S_{n_c}(k_x, k_y)$, where $n_c \in [1, N_c]$ is the coil index and $k_x, k_y$ are k-space coordinates. It is assumed that 2D Cartesian k-space data is arbitrarily sub-sampled along the phase encoding dimension and fully sampled along the readout direction, i.e. the data is sampled for every $k_y \in [1, N]$, $k_x \in K_x$, where $K_x$ designates the set of acquired k-space columns.

*Step I: CORE computation*

This section outlines the first step of the proposed CORE-Deblur method, which implements the CORE technique for computing the convolution image. CORE is a linear parameter-free technique, which is described here shortly; a full mathematical proof can be found in [15].

The aim of CORE is to compute the convolution image $f^{conv}(x, y_0)$ which is defined by,

$$f^{conv}(x, y_0) = f(x, y_0) * g(x) = $$
$$= \sum_{x_o=1}^{N} f(x, y_0) g(x_0 - x) \quad \forall x, y_0 \in [1, N] \quad (1)$$

Each row of $f^{conv}(x, y_0)$ is equivalent to the 1D convolution between the kernel $g(x)$ and the same row in the unknown image $f(x, y)$. We emphasize that CORE computes the convolution image $f^{conv}(x, y_0)$ without performing any explicit convolutions.

Without loss of generality, we now describe how pixel $(x_0, y_0)$ of the convolution image is computed. Let $N_K$ designate the number of acquired k-space columns (for a single coil). To compute $f^{conv}(x, y_0)$, CORE applies a two-step process.

First, it computes a set of weights $W_{y_0}$ that enables the description of the kernel $g(x)$ as a linear combination of the vectors $M_{y_0}$, defined by

$$M_{y_0}(n_{ck}, x) = C(n_c, x, y_0) \cdot \exp(-iK_x(n_k) \cdot x)$$

$\forall n_k \in [1, N_k]$, $n_c \in [1, N_c]$, $y_0 \in [1, N]$; these vectors are modulated versions of the sensitivity maps for column $y_0$. This subspace representation of the kernel $g(x)$ is described by,

$$g(x_0 - x) \equiv \sum_{n_{ck}=1}^{N_c N_k} W_{y_0}(n_{ck}, x_0) \cdot M_{y_0}(n_{ck}, x) \quad (2)$$

$$\forall x, x_0 \in [1, N]$$

where $n_{ck}$ is an index that runs from 1 to $N_c N_k$, i.e. it counts all the acquired k-space columns of all the coils. The weights are computed by solving the ill-posed problem of eq. (2) using the Least Squares approach. Secondly, the obtained weights $W_{y_0}$ are utilized for computing the convolution image from the sub-sampled k-space data using a simple linear combination,

$$f^{conv}(x_0, y_0) = \sum_{n_{ck}=1}^{N_c N_k} W_{y_0}(n_{ck}, x_0) R_{y_0}(n_{ck}) \quad (3)$$

$$\forall x_0, y_0 \in [1, N]$$

where $R_{y_0}(n_{ck}) = F_y^{-1}\{S(n_c, k_x, k_{y_0})\}$, i.e. $R_{y_0}(n_{ck})$ is a vector obtained by applying the 1D Discrete Fourier Transform to the k-space signal obtained by coil $n_c$ at k-space column $k_x$, where this transform is applied along the fully-sampled dimension $y$.

Equations (2) and (3) describe the reconstruction of a single pixel $(x_0, y_0)$ of the convolution image. In practice, there is no need to reconstruct each pixel separately, and CORE reconstructs full rows simultaneously using a simple matrix-form implementation [15].

CORE is a general technique, suitable for various kernel types. In [18] it was implemented in a two-channel process with two wavelet decomposition kernels, hence it produced the wavelet coefficients of $f(x, y)$. In this paper, in contrast, CORE is implemented only once, with a narrow *Gaussian* kernel, hence it produces a slightly blurred version of the target image $f(x, y)$. The blurring is removed in the subsequent CS process, as described below.

*Step II: deblurring by Compressed Sensing*

The second step of CORE-Deblur applies a CS process for removing the inherent blurring in $f^{conv}(x, y)$ and therefore reconstructs $f(x, y)$. This CS process solves the well-established $l1$-minimization problem,

$$\begin{cases} \min & \|\Psi f\|_1 & (a) \\ \text{s.t.} & \mathbf{F}_u \mathbf{F} C_{n_c} f(x, y) = y_{n_c} \quad n_c = 1, ..., N_c \ (b) \end{cases} \quad (4)$$

where $\Psi$ is a sparsifying transform such as the wavelet transform, $\mathbf{F}$ is the Fourier operator, $\mathbf{F}_u$ is a subsampling operator that determines the locations of the acquired k-space data points, and (as defined above) $C_{n_c}$ is the sensitivity map of coil $n_c$. Eq. (4a) represents the desired sparsity of $f(x, y)$ in the $\Psi$ transform domain, and eq. (4b) represents the data fidelity constraint, which is applied to each coil separately. In this work the solution of eq. (4) was obtained by solving its related unconstrained form using the Projection Onto Convex Sets (POCS) approach [17,18].

It is worth emphasizing that CORE-Deblur computes *all* the pixels of $f^{conv}(x, y_0)$, despite the k-space subsampling. CORE-Deblur achieves this by exploiting the mathematical relations between the Fourier transform of sensitivity-weighted data and the convolution operation. Another point worth emphasizing is that in contrast to conventional CS methods, CORE-Deblur does not apply any k-space zero-filling. In conventional CS methods, the $l1$-minimization process described by eq. (4) is commonly initiated by zero-filling the missing k-space data and applying an inverse Fourier transform. However, as is well-known by Shannon's sampling theory, sampling below the Nyquist rate and applying the inverse Fourier transform to the undersampled data results in aliasing. CORE-Deblur *avoids such aliasing* by not performing any zero-filling; instead, it reconstructs the convolution image, which is slightly blurred yet *fully-sampled* in the image domain. As will be shown, initiating the CS process from this image not only avoids aliasing, but also reduces the number of iterations significantly and improves the reconstruction quality.

### III. METHODS

*3.1 Data acquisition*

*Simulations*. reconstructions were performed with simulated data from a realistic analytical brain phantom with eight simulated sensitivity maps using the toolbox developed by Guerquin-Kern et al. [19,20]. The experiments were performed with four different k-space undersampling schemes: (1) periodic undersampling, i.e. equi-spaced acquisition of full columns, (2) varying-period undersampling, in which the columns were sampled using different densities at the k-space center and periphery, (3) random variable-density undersampling, which was implemented here using the SparseMRI toolbox [21], and (4) random uniformly-distributed undersampling.

*In-vivo experiments*: retrospective undersampling experiments were performed with six datasets of brain scans of two healthy volunteers (one female and one male), obtained using a 7T whole-body scanner (Philips Achieva, Best, The Netherlands) equipped with a quadrature transmit head coil and a 32-channel receiver coil array (Nova Medical, Wilmington, MA). The scans were approved by the Leiden University Medical Center ethical committee. Low-resolution scans (gradient echo, TR/TE /FA = 250 ms / 1.9 ms / 40°, 60 x 60 matrix, scan-time 16 s) were first performed, and the coil sensitivity maps were computed from these scans using a simple Sum of Squares (SOS). Next, a series of scans with different contrasts was performed. All sequences used the same field of view (FOV) of 240 x 240 mm$^2$ and an acquisition matrix of 240 x 240. $T_1$-weighted scans used a gradient echo sequence with TR/ TE/ FA = 4.3 ms / 2.05 ms / 7°, TI = 1100 ms, 60 RF excitations per inversion pulse; scan-time (full k-space coverage) was 16s. $T_2$-weighted contrast scans used a turbo spin echo sequence with TR/ TE /FA = 2 s / 70 ms / 90° and 8 refocusing RF pulses; scan-time (full k-space coverage) was 66s. $T_2^*$-weighted contrast scans used a gradient echo sequence with TR/ TE /FA = 250 ms/ 15 ms/ 40°; scan-time (full k-space coverage) was 62s. In all scans, fully-sampled k-space data were acquired, imported to an external computer, and decimated offline.

### 3.2 CORE-Deblur implementation details

To demonstrate the proposed CORE-Deblur method, CORE was implemented using a Gaussian kernel with $\sigma = 0.25$ (unless stated differently). Subsequently, a CS reconstruction that solves eq. (4) was computed using the POCS algorithm, which includes a soft-thresholding operation in the sparsifying transform domain. The POCS algorithm was chosen due to its efficiency, implementation simplicity, lack of constraints on the undersampling pattern and guaranteed convergence [3,17,18]. The sparsifying transform $\Psi$ was chosen to be the Stationary Wavelet Transform (SWT), since its shift-invariant property prevents the production of pseudo-Gibbs artifacts related to thresholding in a decimated wavelet transform domain [22,23]. The CS-POCS method was implemented with a Daubechies-2 wavelet and a threshold of 0.0012. During the CS-POCS iterations, information from the multiple channels was combined using the optimal method of Roemer et al. [24]

$$f = \sum_{i=1}^{Nc} a_i \left( f_i / C_i \right) \text{ where } a_i = C_i^2 / \sum_{j=1}^{Nc} C_j^2 \qquad (5)$$

where $f_i = C_i f$ is the estimated image of coil $i$. The CORE-Deblur algorithm is described schematically in figure 1.

To demonstrate that CORE-Deblur is insensitive to the value of the Gaussiann kernel width, denoted here by $\sigma$, several experiments were performed with in-vivo data using a Gaussian width ranged from $\sigma = 0.25$ to $\sigma = 5$. All other implementation parameters remained the same.

### 3.3 Comparison with CS

To demonstrate the benefits of the CORE computation step, the method was compared to a reconstruction using CS only. The initial guess for this CS method was calculated in the conventional manner, i.e. the sub-sampled k-space data of each coil were zero-filled and inverse Fourier transformed, and then the data from all coils were merged using eq. (5). For a fair comparison, the CS method solved the same reconstruction problem as CORE-Deblur (eq. (4)), and was also implemented using the POCS algorithm, with the same threshold parameter.

### 3.4 Reconstruction quality assessment

The reconstruction quality of both methods was measured using the Normalized Root Mean Square Error (NRMSE) in relation to the corresponding gold standard image. The latter was computed by applying the inverse Fourier transform to the fully sampled

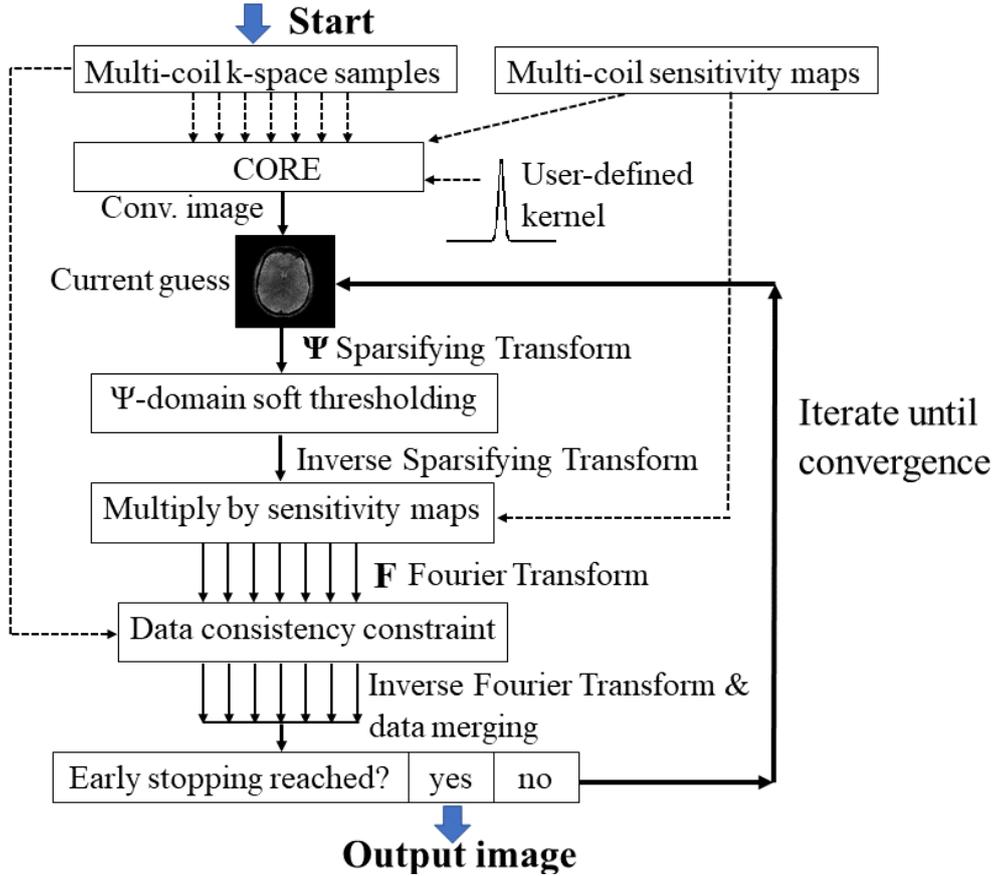

Figure 1. Schematic depiction of the proposed CORE-Deblur method. Dashed lines designate inputs.

coil-specific k-space data, with a subsequent coil merging step using eq. (5).

All computations were carried out in Matlab R2017a (The MathWorks Inc., Natick, MA, 2000) on an HP-Spectre x360 computer, with Intel CORE™ core i7 (7$^{th}$ generation) and 16 GB of RAM.

## IV. RESULTS

### 4.1 Numerical brain phantom

Figure 2 shows reconstructions of the numerical brain phantom [19] obtained from the four different k-space subsampling schemes described above, which were all implemented with a high reduction factor of R=10. In all cases, the CS method produced reconstructions with severe aliasing, which reflected the undersampling patterns; these artifacts are noticeable in reconstructions obtained even after 2500 iterations. In contrast, and despite the high undersampling rate, CORE-Deblur produced high-quality reconstructions, as quantitatively reflected by low NRMSEs, and required only 10 iterations. These results indicate that CORE-Deblur is robust with respect to the k-space subsampling pattern and enables a wide range of subsampling schemes. In contrast, many CS-PI methods often adopt the variable-density undersampling pattern [3,25], to avoid artifacts created by periodic sampling.

### 4.2 In-vivo T2* imaging

Reconstruction results obtained from in-vivo T2* scan data with periodic subsampling and an acceleration factor of R=5 are presented in Figure 3a. The results demonstrate that while the CS initial guess was of low quality, CORE-Deblur produced an initial guess very similar to the gold standard. Moreover, CS required 90 iterations, whereas CORE-Deblur was stopped after 11 iterations only; despite this 10-fold difference in the number of iterations, CORE-Deblur obtained a lower reconstruction error. As can be observed (Figure 3b), the CS process initiated from a high NRMSE and reduced slowly, while the CORE-Deblur process initiated from a much lower error and reduced quickly. These results suggest that CS would not benefit from an early stopping criterion due to its high initial error.

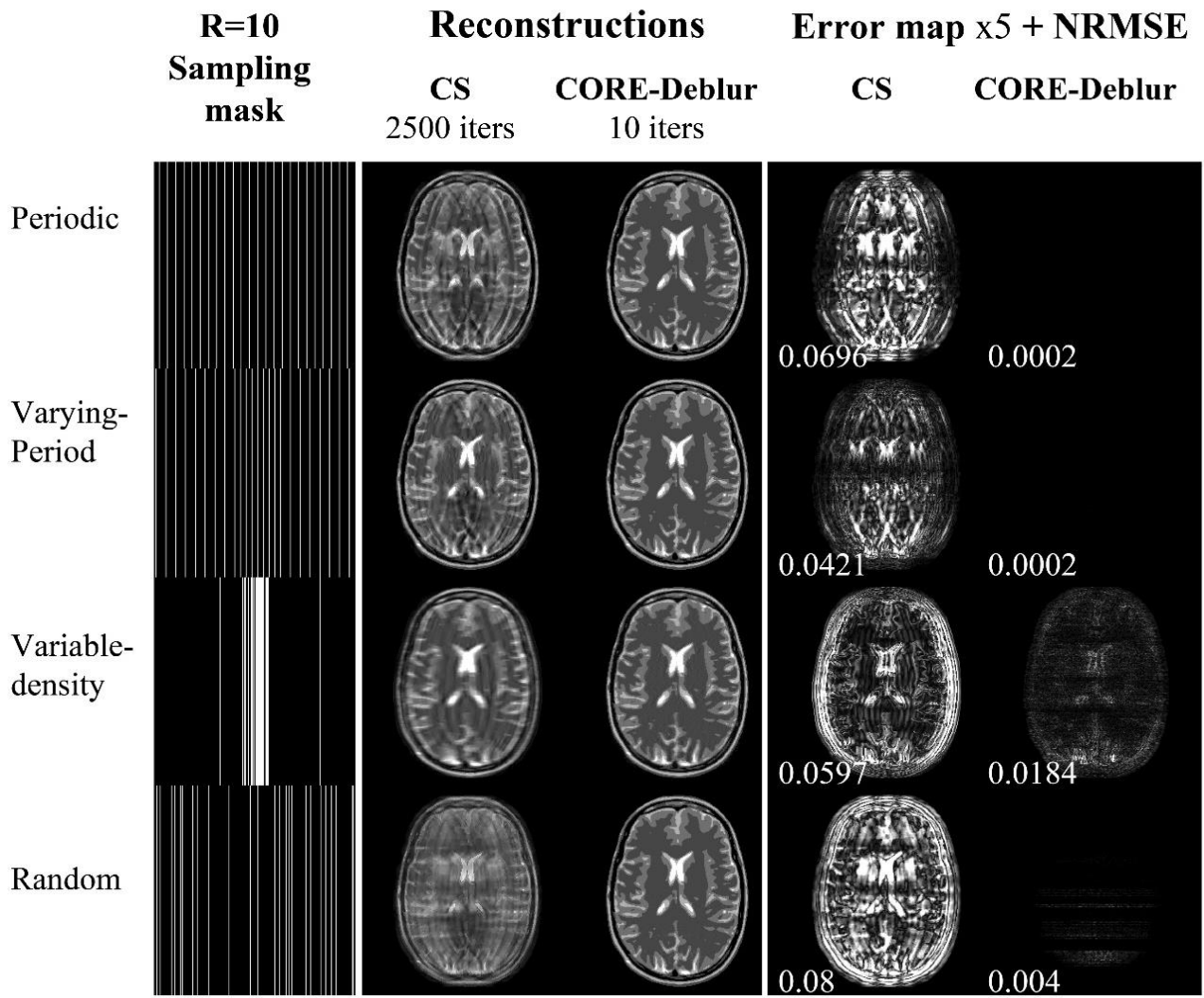

Figure 2. Reconstructions of an eight-coil simulated brain phantom. K-space was subsampled with a reduction factor of $R=10$ using four different undersampling schemes (left column). This figure highlights that the proposed CORE-Deblur method produces an alias-free reconstruction after 10 iterations only, whereas the conventional CS method does not remove the aliasing after 2500 iterations.

*4.3 In-vivo T1 imaging*

Reconstructed images for an in-vivo T1 coronal scan with R=5 are depicted in Figure 4a. The CS method produced an initial guess with aliasing (top row) and converged to a high-quality reconstruction after 99 iterations. CORE- Deblur, on the other hand, produced an initial guess that was noisy but preserved structures within the image (top row), and required only 6 iterations. Furthermore, the final CORE-Deblur reconstruction had a lower NRMSE than the one obtained by CS after 99 iterations. The error graphs (Figure 4b) show that CORE-Deblur also exhibited a rapid NRMSE decrease, whereas the CS method showed a much slower decrease.

*4.4 CORE-Deblur with various kernel widths*

To demonstrate that CORE-Deblur is robust with respect to the Gaussian kernel width, experiments were performed with various σ values in the range of 0.25 to 5. The results (Figure 5) demonstrate that CORE-Deblur obtained high-quality reconstructions after 10 iterations in all cases, even for a highly blurred (σ=5) initial guess (Figure 5, right column). Over the entire range, the NRMSE did not exceed 0.03 and the reconstructions did not exhibit any noticeable aliasing. These results suggest that CORE-Deblur is insensitive to the kernel width.

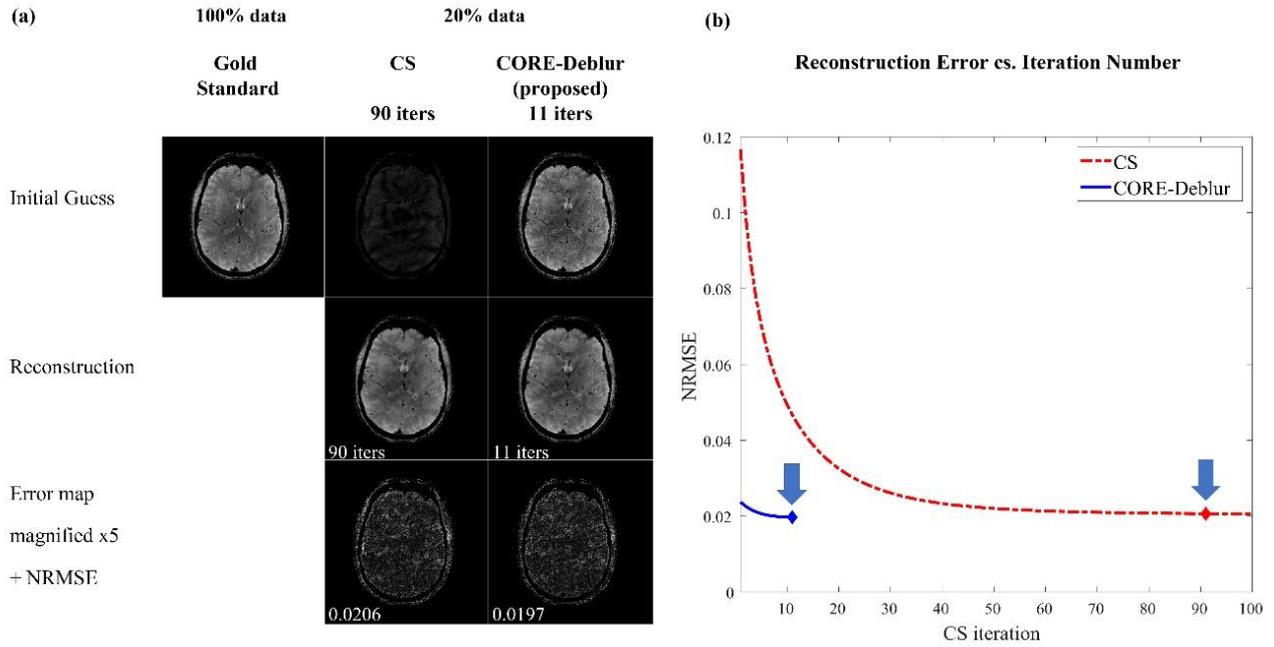

Figure 3. Reconstructions from in-vivo data, subsampled with 5-fold acceleration. (a) Comparison of the gold standard image (left column) with reconstructions obtained by a conventional CS process (middle column) and the proposed CORE-Deblur method (right column). (b) Reconstruction error vs. iteration number. This figure highlights that the initial guess of CORE-Deblur is much closer to the gold standard than the CS one, and that CORE-Deblur requires much fewer iterations for producing an accurate reconstruction.

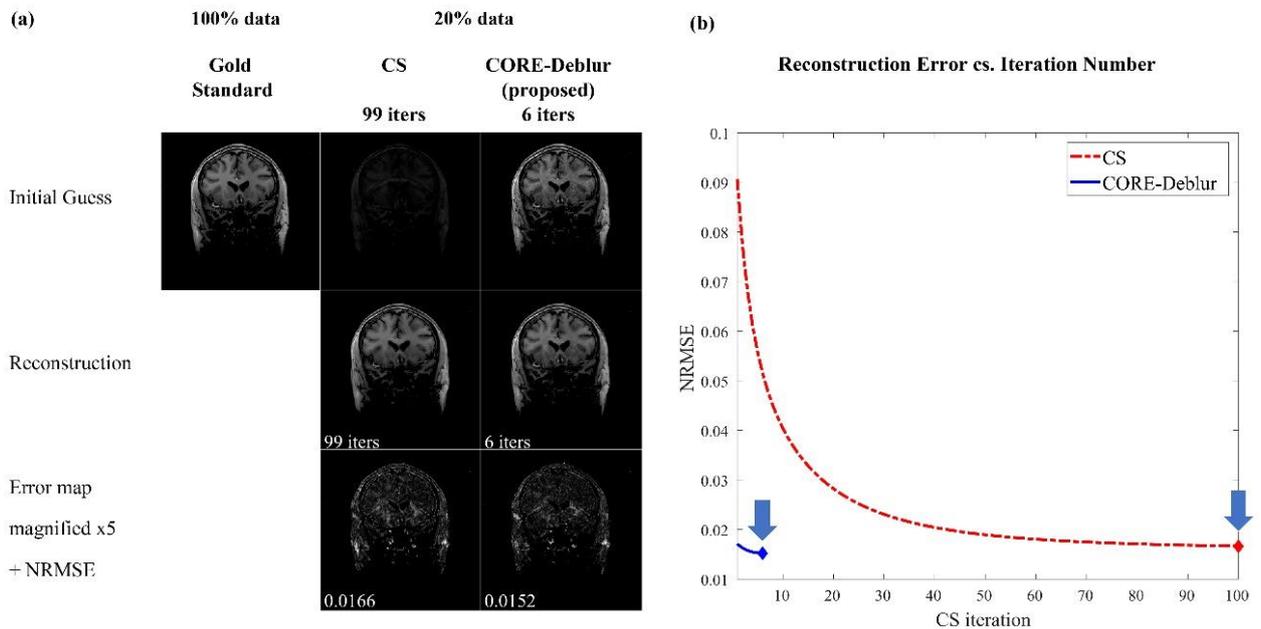

Figure 4. Reconstructions from in-vivo 5-fold subsampled k-space data. Note that CORE-Deblur produces a high-quality reconstruction after 6 iterations only, whereas the CS method requires 99 iterations for convergence.

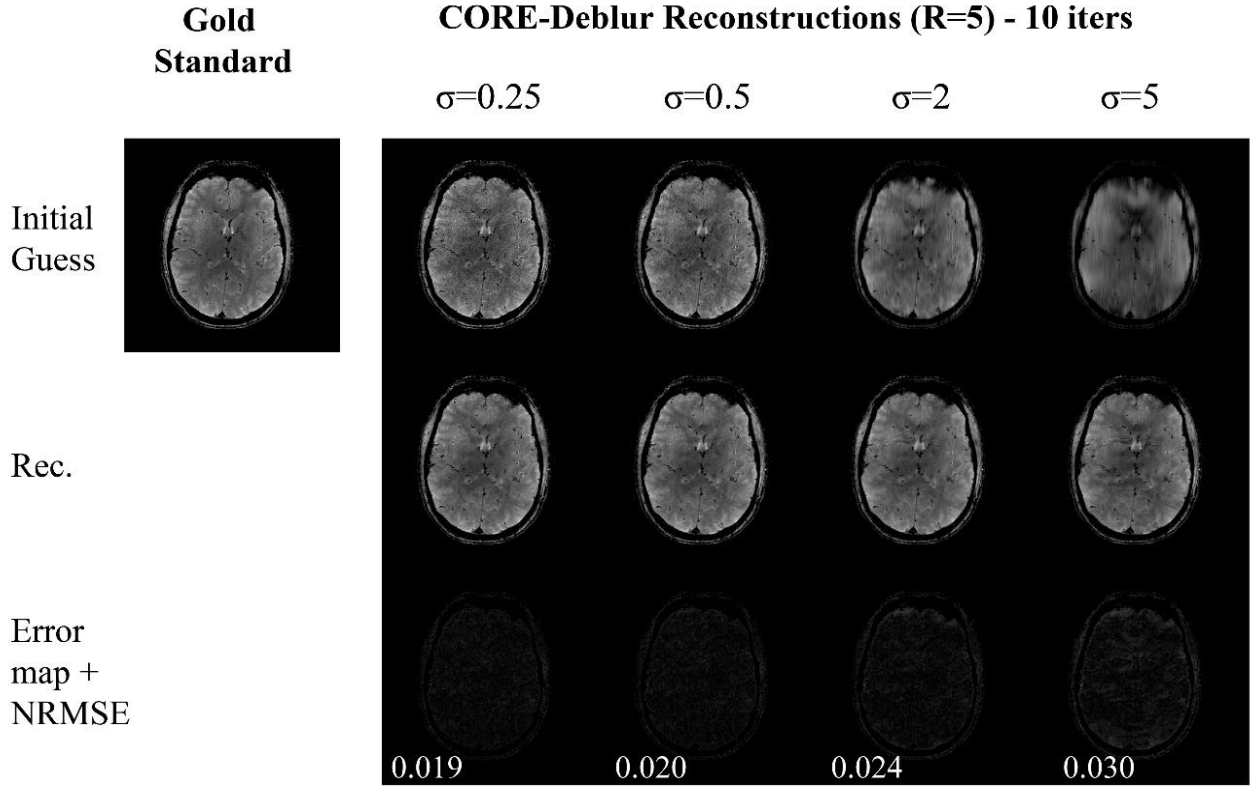

Figure 5. CORE-Deblur reconstructions from subsampled data (R=5) for Gaussian kernels with different widths. Note that CORE-Deblur yielded high quality reconstructions within 10 iterations in all cases, even when the initial guess was very blurred (right column).

*4.5 Statistical analysis of in-vivo scans*

Figure 6 and Table 1 describe the average results obtained for six in-vivo scans (three T1 scans, two T2* scans and one T2 scan), in which k-space was undersampled periodically with R=5. The curves in figure 6 depict the average and standard deviation of the NRMSE vs. CS iteration number. As can be observed, the average initial error of the CS method was much higher than that of CORE-Deblur; additionally, the CS average error decreased moderately as a function of the number of iterations, whereas the CORE-Deblur average error converged much more rapidly. Additionally, the results in Table 1 demonstrate that the CS method would not benefit from early stopping after 10 iterations, since at that point it exhibits high errors. Moreover, these results demonstrate that CORE-Deblur - with 10 iterations only - produces a lower reconstruction error than that obtained by CS after 100 iterations. These results show that CORE-Deblur benefits from the high quality of its initial guess and from the implicit regularization that is obtained by early stopping of the iterative process, and that it enables reducing the CS iterations number by a factor of 10.

## V. DISCUSSION

This work describes a new hybrid method for image reconstruction from multi-coil subsampled k-space data. CORE-Deblur first uses the CORE technique for reconstructing a blurred convolution image, and then implements a short CS process (with few iterations) for removing the convolution-related blurring and reconstructing the final image. Simulations and retrospective experiments with in-vivo data demonstrated that CORE-Deblur: (1) enables high-quality reconstruction from arbitrarily subsampled k-space data, with flexible subsampling schemes, (2) exhibits robustness with respect to the initial kernel width, and (3) reduces the number of CS iterations by 10-fold compared with a conventional CS reconstruction. CORE-Deblur therefore addresses two of the limitations of CS-PI methods: the need for time-consuming random subsampling trajectories, and lengthy iterative computations.

*Relation to previous work*

Generally, $l1$-minimization has been proposed as a powerful mechanism in a variety of image deblurring methods in the image-processing literature [26–28]; recently, such methods were suggested in the context of dictionary-based image deblurring [27,29,30]. However, these methods have all addressed the classical image deblurring problem, in which a latent image needs to be recovered from noisy measurements of its blurred version; in this problem, the measurements are obtained in the image domain (i.e. not in k-space). To the best of our knowledge, this is the first work that proposes an $l1$-minimization *deblurring* process as part of an MR image *reconstruction* method. Furthermore, while different general-purpose image-deblurring methods have exploited early stopping of iterative algorithms as a source of implicit regularization [31–33], this work uniquely utilizes this mechanism for MR image reconstruction.

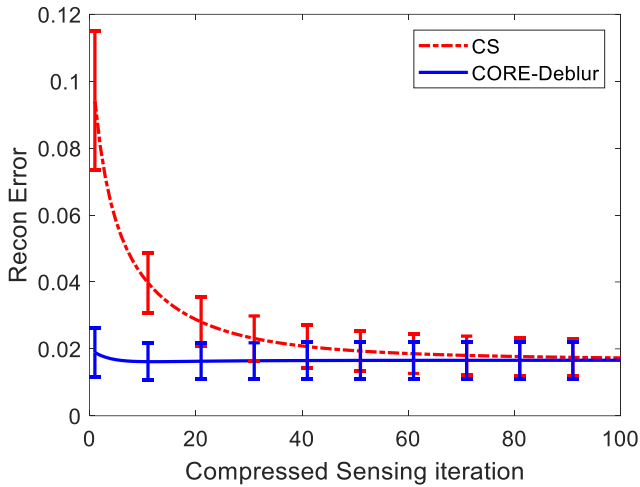

Figure 6. Average NRMSE and its standard deviation for reconstructions obtained using 5-fold undersampled k-space data from six in-vivo scans. Note that CORE-Deblur initiates from a lower NRMSE than CS and converges much faster.

|  | CS |  | CORE-Deblur |
| --- | --- | --- | --- |
| Dataset | 10 Iters. | 100 Iters | 10 Iters. |
| 1 | 0.050 | 0.021 | 0.020 |
| 2 | 0.056 | 0.027 | 0.026 |
| 3 | 0.029 | 0.010 | 0.009 |
| 4 | 0.044 | 0.018 | 0.016 |
| 5 | 0.032 | 0.013 | 0.011 |
| 6 | 0.040 | 0.017 | 0.015 |
| Av. Error | **0.040** | **0.020** | **0.016** |

Table 1. Reconstruction Normalized Root Mean Square Errors (NRMSEs) for CS and CORE-Deblur in experiments with in-vivo brain datasets.

Image deblurring methods often solve a *blind* deblurring problem, hence they require computationally-expensive algorithms for estimating the unknown deblurring kernel [29,30]. In contrast, CORE-Deblur has the benefit of a *known* blurring kernel since this kernel is defined by the user in the first step of this method. CORE-Deblur hence solves a *non-blind deconvolution problem*. This problem is traditionally solved using deconvolution algorithms such as the Wiener Filter [34] or the Lucy-Richardson algorithm [35]. However, these methods are known to produce reconstructions with ringing effects. Since CORE-Deblur avoids explicit deconvolution operations, it avoids such undesired effects; the in-vivo results presented here demonstrate that CORE-Deblur produces reconstructions without such artifacts.

The proposed method implements the CORE technique, which is a mathematical method that enables computation of the convolution between an unknown MR image and a known user-defined kernel. CORE was originally introduced in [15] as part of the CORE-PI method; in that work, CORE was implemented in a two-channel process, with two wavelet kernels that represent the low-pass and high-pass decomposition kernels of the Stationary Wavelet Transform (SWT). Together, these operations produce the SWT coefficients of the target image, and the image is subsequently reconstructed using the Inverse SWT. In the present work, in contrast, CORE is applied only once, with a Gaussian kernel; it produces the convolution image, which is a blurry version of the target image. As demonstrated in the results, the target image can be subsequently reconstructed using a CS process. CORE-Deblur is hence different from the method proposed in [18] in several aspects: CORE-Deblur employs a single-channel process, utilizes a different kernel, and implements a short iterative reconstruction (rather than the Inverse SWT) for final image recovery.

*k-Space subsampling flexibility*

The CS theory fundamentally requires random k-space subsampling to achieve incoherent aliasing in the reconstruction [36], since ordered subsampling produces periodic artifacts that cannot be removed by a CS reconstruction process (see for example figure 3). Methods integrating the CS and PI frameworks therefore often exhibit degraded performance for regular subsampling schemes. In contrast, the results presented here demonstrate that CORE-Deblur enables various subsampling schemes, both regular and random. This flexibility is enabled because CORE-Deblur *avoids* the k-space zero filling step that is common to many CS methods, hence it *prevents any potential periodic aliasing*. CORE-Deblur is therefore robust with respect to the undersampling scheme.

*Practical limitations*

A possible limitation of CORE-Deblur is its requirement for estimating the coil sensitivity maps prior to image reconstruction; this may lead to increased noise in comparison to auto-calibration methods or calibrationless methods [37–39]. However, CORE-Deblur does not exhibit inhomogeneity intensity artifacts that are related to auto-calibration [40], and it uses significantly simpler computations than auto-calibration methods. Possible extensions of CORE-Deblur may either attempt to improve the sensitivity maps estimation during the CS reconstruction [25,40] or incorporate an auto-calibration type process. Another limitation of CORE-Deblur is that it is currently suitable only for 2D Cartesian data; the method may be expanded into 3D Cartesian imaging, but this is beyond the scope of the current work.

## VI. CONCLUSIONS

In this work, the CORE-Deblur method is introduced for image reconstruction from multi-coil subsampled k-space dat. This work also proposes the novel concept of image reconstruction by applying CS for image deblurring. Experiments with a simulated brain phantom and in-vivo 7T data demonstrated that CORE-Deblur enables: (i) high-quality reconstructions, (ii) reduction of the number of required CS iterations by 10-fold compared to CS reconstruction alone, (iii) robustness with regard to the initial kernel width, and (iv) compatibility with various k-space undersampling schemes, ranging from regular to random.

## ACKNOWLEDGEMENT

The authors thank Michael Elad for his useful comments and the Technion Society of the Netherlands for their support. A. G. W. was supported by ERC ADG 670729 NOMA MRI.